# Absence of Debye Sheaths Due to Secondary Electron Emission

M. D. Campanell, A. Khrabrov and I. D. Kaganovich
*Princeton Plasma Physics Laboratory, Princeton University, Princeton, New Jersey 08543, USA*

A bounded plasma where the electrons impacting the walls produce more than one secondary on average is studied via particle-in-cell simulation. It is found that no classical Debye sheath or space-charge limited sheath exists. Ions are not drawn to the walls and electrons are not repelled. Hence the plasma electrons travel unobstructed to the walls, causing extreme particle and energy fluxes. Each wall has a *positive* charge, forming a small potential barrier or "inverse sheath" that pulls some secondaries back to the wall to maintain the zero current condition.

Any unbiased material in contact with a plasma must draw zero net current in equilibrium. Typically, the thermal velocity of electrons is much larger than that of ions. So the material charges negatively. A strong electric field called the Debye sheath [1] forms at the boundary, accelerating ions to the surface while reflecting enough approaching electrons to keep the two fluxes equal. Sheath theory is essential for studying plasma-wall interaction, setting boundary conditions in fluid simulation codes [2,3] and measuring plasma properties with Langmuir probes [4].

If the bulk plasma has a Maxwellian electron velocity distribution function (EVDF), the electron flux to the surface can be written as $\Gamma_e = \Gamma_{e,0}\exp(-e\Phi/T_e)$ [5] where $\Gamma_{e,0} = n(T_e/2\pi m_e)^{1/2}$ is the thermal flux that would strike a surface *without* the insulating sheath barrier. Here $\Phi$ is the sheath potential magnitude, $n$ is the plasma density and $T_e$ is the electron temperature. The ion flux $\Gamma_i$ given by the Bohm criterion [1,6] is independent of $\Phi$. So $\Phi$ is easily computed by solving $\Gamma_e = \Gamma_i$. There are variations [3,4] of sheath theory accounting for non-Maxwellian EVDF's due to magnetic fields, low collisionality, applied beams, *etc*. But the qualitative features are similar. The sheath forms a potential barrier $\Phi$ of magnitude needed to maintain zero current.

Bombardment from plasma electrons may eject electrons from a material. For most materials in the energy range of interest ($\varepsilon < \sim 400$ eV), $\gamma(\varepsilon)$, the average number of "secondaries" produced by an incident electron, increases with impact energy $\varepsilon$ [7]; (the dependence on impact angle is weaker and not crucial in this discussion). Secondary electron emission (SEE) alters the current balance. Let $\gamma_{net} \equiv \Gamma_{out}/\Gamma_{in}$ denote the ratio of emitted flux to incident electron flux at a wall. The zero current condition becomes,

$$\Gamma_e = \Gamma_{in} - \Gamma_{out} = \Gamma_{in}(1-\gamma_{net}) = \Gamma_i. \qquad (1)$$

$\gamma_{net}$ depends on the distribution of impact energies and generally increases with $T_e$. As $\gamma_{net}$ increases, more electrons must reach the wall to balance the fixed ion flux, so $\Phi$ decreases, allowing $\Gamma_{in}$ in (1) to increase. At high temperatures as $\gamma_{net} \to 1$, the influx increases rapidly because $\Gamma_{in} = \Gamma_i/(1-\gamma_{net})$. Before $\gamma_{net}$ can reach unity, the emission $\Gamma_{out} = \gamma_{net}\Gamma_i/(1-\gamma_{net})$ becomes intense enough that the negative charge formed by secondaries at the interface creates a potential barrier that reflects some cold secondaries back to the wall [8]. In principle, this allows zero current to be maintained even if the emission induced by hot plasma electrons exceeds unity. The *net* emission $\gamma_{net}$ saturates to a critical value $\gamma_{cr} < 1$ so that the ion flux in (1) can still be balanced. The "space-charge limited" (SCL) sheath is usually assumed to form under very strong emission in tokamaks [2,9], Hall thrusters [10], emissive probes [3,11] and general plasma-wall systems [4,8].

All theories invoking the SCL sheath rely on a sheath structure existing *a priori* as the SEE intensity increases beyond the threshold for saturation. For instance, the original Hobbs-Wesson paper [8] assumes ions "arrive at the sheath edge" with a velocity related to the Bohm criterion. Poisson's equation is then solved with charge densities written in terms of the *negative* potential assumed in the sheath. However, suppose a material is suddenly inserted into a hot plasma. The initial rush of electrons with $\gamma(\varepsilon) > 1$ will cause *reduction* of electrons on the surface and ions would then be repelled from the surface. The assumptions [1,4,6] inherent in deriving the Bohm criterion are not satisfied, *e.g.* that the wall potential is negative with respect to the plasma and that ions are drawn to the wall. Morozov and Savel'ev [5] have shown that for a plasma-wall-SEE system with a Maxwellian EVDF at infinity, at high temperatures there are potential profile solutions in which the wall potential is indeed positive with respect to the plasma. Overall, it is unclear whether a sheath could form in the first place if SEE is very strong.

In this Letter, we study directly by particle simulation a plasma in which electrons impacting the walls on average have $<\gamma(\varepsilon)> > 1$. This situation may naturally arise in a Hall thruster (HT) when the E×B drift velocity $V_D$ is large. We simulate such a plasma using EDIPIC code and show the behavior is unexplained by familiar theories. In particular, there is no classical sheath or SCL sheath. Electrons travel unimpeded to the walls. The plasma in this new regime is dramatically different than in past EDIPIC HT simulations [12,13,14,15,16] with smaller drift energy.

EDIPIC (electrostatic direct implicit particle-in-cell code) simulates a planar E×B xenon plasma bounded by floating walls made of boron-nitride ceramics (B.N.C.), see Fig. 1(a). Details on the numerical algorithms, verification and past results are provided in Ref. 16. Both the plasma and sheath regions are resolved. The applied fields $E_z$ and $B_x$ are uniform. Both ions and electrons are treated as particles. The plasma is given an initial density $n_0$ and allowed to evolve.

Particle dynamics are governed by the plasma's self-generated field $E_x(x)$ and the E×B drift motion from the background fields. For electrons, the neutral gas density $n_a$ determines the frequency of elastic collisions $\nu_{en}$. Coulomb collisions are implemented with a Langevin model, but can usually be neglected as they only weakly affect HT's [17]. Turbulent collisions of frequency $\nu_{turb}$ effectively simulate anomalous conductivity by scattering the y-z component of the velocity vector [18]. Each scatter leads to displacement along $E_z$ and a corresponding energy gain parallel to the *walls* on average of $<\Delta W_{//}> = m_e V_D^2$.

Past simulations modeling the PPPL HT have found that in the low collisionality regime anticipated in experiments, the bulk plasma EVDF is anisotropic and strongly depleted in the loss cone [12]. In contrast to collisional HT regimes where the SEE thermalizes in the plasma [10], the emitted electrons form beams that cross the plasma and strike the other wall. The particle flux at each wall consists of collision-ejected electrons (CEE's) scattered into the loss cone by impacts with neutrals, "beam" electrons from the other wall and ions. $\Gamma_i$ is given by the Bohm criterion in terms of the effective electron temperature normal to the walls, $T_x$ [12]; $\Gamma_i \approx (n/2)(T_x/m_i)^{1/2}$. In a quasisteady state, the zero current condition applies. By symmetry, the two beams are equal and opposite. So at each wall, the incoming beam and outgoing SEE are equal. Eq. (1) becomes $\Gamma_e = (\Gamma_{CE} + \Gamma_b) - \Gamma_b = \Gamma_{CE} = \Gamma_i$. Also, the SEE produced by $\Gamma_{in}$ must yield the outgoing beam $\Gamma_b$. That is, $\gamma_{CE}\Gamma_{CE} + \gamma_b\Gamma_b = \Gamma_b$. We obtain,

$$\Gamma_{CE} \approx \Gamma_i, \quad (2)$$

$$\Gamma_b \approx \frac{\gamma_{CE}\Gamma_{CE}}{(1-\gamma_b)}, \quad (3)$$

$$\gamma_{net} \approx \frac{\gamma_{CE}}{(1+\gamma_{CE}-\gamma_b)}, \quad (4)$$

where $\gamma_b$ and $\gamma_{CE}$ are the partial SEE coefficients. (e.g., $\gamma_{CE} \equiv$ ratio of secondary flux produced by CEE's to $\Gamma_{CE}$). It has been found [12] that a classical non-SCL sheath forms even if $\gamma_{CE}$ is well above unity because as long as $\gamma_b < 1$, $\gamma_{net} < 1$ also via (4). Past simulations typically used $B_x = 100G$ and $E_z = 50\text{-}200V/cm$ to model experimental conditions. For E = 200V/cm, it was found that $\gamma_b$ approaches unity (~0.92-0.95). This is because the drift energy gained by cold emitted secondaries crossing the plasma can range up to $2m_e V_D^2 = 45eV$, so the beam energy can approach the $\gamma(\varepsilon) = 1$ threshold for B.N.C., where $\gamma(\varepsilon) \approx 0.17\varepsilon^{1/2}$ ($\varepsilon$ in eV) [19].

If $E_z$ is increased further, the physics fundamentally changes. Simulation A ("Sim. A") with $E_z = 200V/cm$, $B_x = 100G$, $n_a = 10^{12}$ cm$^{-3}$, $n_0 = 10^{11}$ cm$^{-3}$, $\nu_{turb} = 1.4 \times 10^6 s^{-1}$ and H = 2.5cm features the familiar behavior discussed previously. We compare it to Simulation B with all conditions equal except $E_z = 250V/cm$ and $\nu_{turb} = 2.8 \times 10^6 s^{-1}$. One may expect the plasma in Sim. B to be hotter with a larger sheath potential, but otherwise similar to Sim. A. Fig. 1 shows the electrostatic potential function $\Phi(x)$ in both runs. Sim. A exhibits a nearly symmetric potential well of amplitude $\Phi \approx 21V$ with well-defined sheaths near the walls. (The asymmetry is from fluctuations due to two-stream instability that arise when the SEE beams are intense [20]). However, Sim. B has no apparent sheath structure. Two-stream fluctuations of a few Volts dominate the shape of $\Phi(x)$.

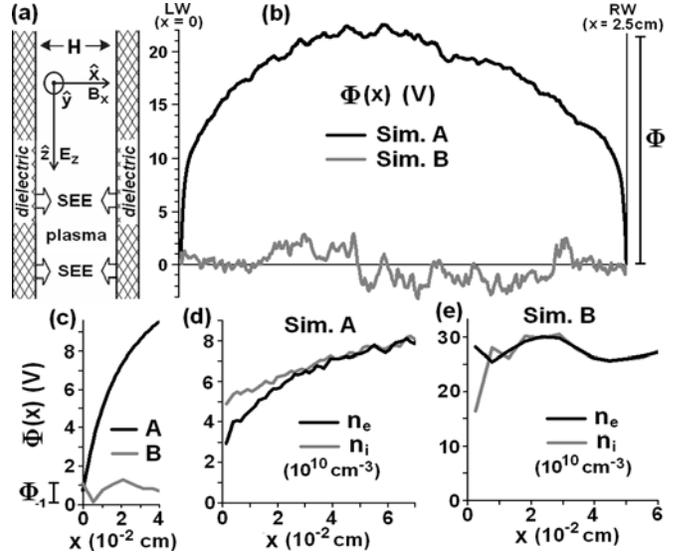

FIG. 1. (a) Simulation model. (b) $\Phi(x)$. (c) $\Phi(x)$ near the left wall (LW). Electron and ion densities near the LW in Sim. A (d) and Sim. B (e). Snapshots (b-e) represent t = 10ms in both runs.

The unusual behavior in Sim. B is due to the SEE. The flux components and partial SEE coefficients are listed in Table 1. In Sim. A, a classical sheath appears because $\gamma_b < 1$ and thus $\gamma_{net} < 1$. Eqs. (2-4) apply. In Sim. B, the E×B drift energy is ~50% larger and $\gamma_b$ actually exceeds unity. Eq. (4) suggests a classical sheath cannot exist because $\gamma_{net}$ too would exceed unity and the ion flux in (1) could not be balanced. Also, with $\gamma_b > 1$, the SEE beams would multiply at each flight between the walls and grow perpetually.

| Simulation | A | B | | A | B |
|---|---|---|---|---|---|
| $\gamma_b$ | 0.94 | 1.22 | $\Gamma_b$ | 78.7 | 1104 |
| $\gamma_{CE}$ | 1.75 | 1.28 | $\Gamma_{CE}$ | 3.21 | 18.2 |
| $\gamma_{net}$ | 0.96 | 1 | $\Gamma_o$ | N/A | 248 |
| $<W_x>$ (eV) | 5 | 2.5 | $\Gamma_{in}$ | 81.9 | 1370 |
| $<W_{//}>$ (eV) | 89 | 46 | $\Gamma_i$ | 2.51 | 0.63 |
| $<V_z>$ (km/s) | -6.5 | -50 | | | |

Table 1. Key parameters at t = 10ms in both runs, after quasisteady state was reached. Fluxes are at the LW in units of $10^7$ cm$^{-2}$ns$^{-1}$.

Closer study of the new regime reveals each wall acquires a slight *positive* charge, as is reasonable to expect if most incident electrons have $\gamma(\varepsilon) > 1$. Ions are repelled away from the wall and the net space charge near the interface is negative, see Fig. 1(e). Therefore, at all times it is found that $\Phi(x)$ decreases from the wall outward, see Fig 1(c). (These features are all opposite to Sim. A.) The small potential barrier at the interface pulls some of the SEE back to the wall. Note that while the barrier amplitude $\Phi_{-1} \approx 1V$ appears trivial relative to the large fluctuations throughout the plasma gap, only the structure of $\Phi(x)$ near the wall affects secondaries near the wall. 1V is sufficient to pull back a substantial fraction of cold secondaries, which are emitted with a thermal distribution corresponding to $T_{emit} = 2eV$. Thus, this "inverse sheath" prevents unbounded charge flow between the plasma and wall, as does a classical sheath. But in contrast, the latter works by reflecting a large portion of *hot* plasma electrons *approaching* the wall, requiring a much larger barrier amplitude $e\Phi \sim T_e$, as in Sim. A.

The zero current condition is still maintained in the inverse sheath regime. Consider the time evolution of the fluxes in Fig. 2. In Sim. B, there are three components of $\Gamma_{in}$; CEE's ($\Gamma_{CE}$), secondaries from the *opposite* wall ($\Gamma_b$) and "other" electrons ($\Gamma_o$). $\Gamma_o$ consists of secondaries pulled back to the wall by the inverse sheath. These electrons are cold and induce no SEE ($\gamma_o = 0$). This is why $\gamma_{net}$ does not exceed unity even though $\gamma_b, \gamma_{CE} > 1$ in Fig. 2. In fact, $\gamma_{net}$ appears to be exactly unity. To see why, first consider the ion flux. In Sim. A, ions are accelerated by the sheath to the wall, forming a substantial flux $\Gamma_i$. The sheath limits $\Gamma_{CE}$ to maintain (2) approximately, see Table 1. In Sim. B, because there are no sheaths, ions are not drawn to the wall and the Bohm criterion does not apply. $\Gamma_i$ is merely 3% of $\Gamma_{CE}$. ($\Gamma_i$ is nonzero because with $T_{ion}$ set to 1eV in the simulation, some ions have sufficient *thermal* energy to overcome the barrier $\Phi_{-1}$ and reach the wall.) With very small $\Gamma_i$, the net electron flux $\Gamma_e$ must be near zero in Sim. B if the current (1) is to be balanced. Hence $\gamma_{net} = 0.9994 \approx 1$ at $t = 10ms$.

Although zero current is necessary for equilibrium in a plasma-wall system, equilibrium itself is not necessary. The plasma in Sim. B exhibits waves and instabilities, fluctuating strongly, as is evident by the irregular shape of $\Phi(x)$ in Fig. 1(b). It might be expected that with no classical sheaths, the current may fluctuate too. But zero current is maintained by the inverse sheath in a *stable* manner at all times, not just on average. In Fig. 2, $\gamma_{net} = 1$ and never varies, despite $\Gamma_b$ and $\Gamma_o$ rapidly fluctuating. Notice that the fluctuations of $\Gamma_o$ closely follow fluctuations of $\Gamma_b$. It turns out the following relation is maintained,

$$\Gamma_b(1-\gamma_b) + \Gamma_o \approx 0. \quad (5)$$

That is, the number of pulled-back secondaries always self-adjusts to make $\gamma_{net} = 1$. Since $\gamma_b \approx 1.2$ roughly in Fig. 2, $\Gamma_o(t) \approx 0.2\Gamma_b(t)$. Eq. (5) is just the equilibrium current equation (1) with $\Gamma_{in} = \Gamma_b + \Gamma_o$, $\Gamma_{out} = \gamma_b\Gamma_b$, and the much smaller terms $\Gamma_i$, $\Gamma_{CE}(1-\gamma_{CE})$ neglected.

The reason for stability is qualitatively simple. If a perturbation in $\Gamma_b$ or $\gamma_b$ causes the floating wall's charge to increase (become more positive), then $\Phi_{-1}$ increases in magnitude. A larger fraction of the emitted secondaries is pulled back to the wall, causing the wall charge to decrease, canceling the perturbation. Hence the inverse sheaths are stable in a current-voltage sense. This is in contrast to classical sheaths in the HT system which were found to become unstable, leading to oscillations of the sheath potential and net current [15, 16]. Interestingly, $\gamma_{net}$ was observed to briefly exceed unity during these instabilities and then the sheaths became SCL. Our introductory discussion suggests that SEE saturation occurred via SCL sheath in those references because a classical sheath with $\gamma_{net} < 1$ and a negative wall potential *already existed* before instability, (i.e. before $\gamma_{net}$ jumped above unity). In the new regime, the SEE yield from plasma electrons also exceeds unity, but since there is no sheath in the first place, SEE saturation occurs via inverse sheath.

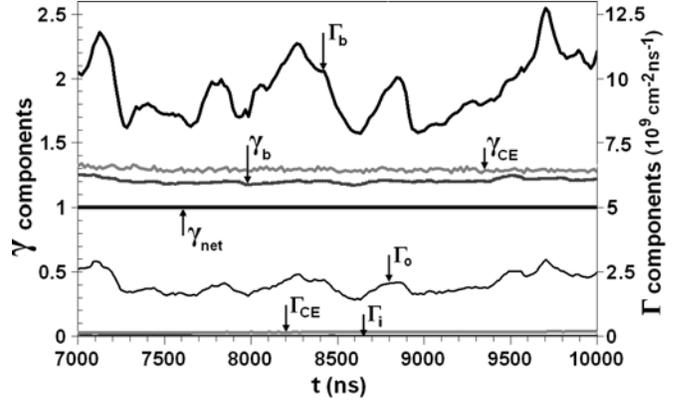

Fig. 2. Temporal evolution of the particle fluxes and partial SEE coefficients at the LW in Sim. B.

The disappearance of the Debye sheath has important implications. Sheaths "insulate" the walls from a plasma by reflecting most approaching electrons. Fig. 3 shows the $EV_xDF$ (the EVDF integrated over $V_y$ and $V_z$) in both simulations. In general in low collisionality, the EVDF is nonlocal [21]. In Sim. A with classical sheaths, bulk plasma electrons throughout the interior of the plasma volume with $\frac{1}{2}m_eV_x^2 < e\Phi$ are trapped and oscillate in the potential well. They cannot hit the wall unless they have large $V_{//}$ and get scattered into the loss cone ($W_x > e\Phi$) by a neutral collision. Because collisionality is low, replenishment of the loss cone is weak and there is a sharp cutoff in the bulk $EV_xDF$ at $V_x = \pm V_{cutoff} \equiv (2e\Phi/m_e)^{1/2}$. Secondaries form a small hump beyond the cutoff velocity. Overall, the walls are protected from most electrons in the system. Electrons gain energy $W_{//}$ parallel to the walls by drift rotation and turbulent collisions. The equilibrium temperature depends on a balance between the collisional heating and collisional losses; $T_{//}$ scales as $E_z^2\nu_{turb}/\nu_{en}$ [12]. So one would expect the plasma in Sim. B to be ~3 times as hot as in Sim. A.

However, because the sheaths vanish in Sim. B, every aspect of the plasma is different. Electrons travel freely to the walls, so all electrons are "secondaries" recently emitted from a wall. There is no sharp cutoff in the $EV_xDF$, but because $V_x$ comes only from the small thermal velocity of emission, the average kinetic x-energy $<W_x>$ is less for Sim. B than Sim. A in Fig. 3. Also, because no electrons are trapped, most will reach the wall before suffering any collisions that increase $W_{//}$. Therefore, $\langle W_{//}\rangle$ is also much smaller. EDIPIC diagnostics record the average kinetic energy of all electrons in the plasma. At t = 10ms in both simulations, the losses and heating are in balance. Sim. A has $\langle W_x\rangle$ = 5eV and $\langle W_{//}\rangle$ = 89eV. Sim. B has $\langle W_x\rangle$ = 2.5eV and $\langle W_{//}\rangle$ = 46eV. The most important feature of Sim. B is that the particle/energy fluxes to the walls are enormous. With no sheaths and $\gamma_{net}$ = 1, *all* electrons in the system can be thought of as traveling back and forth from wall to wall repeatedly. Because of this, $\Gamma_{in}$ is 17 times larger in Sim. B compared to Sim. A, see Table 1. The secondaries, though emitted cold, will displace along $E_z$ and gain drift energy before impacting the other wall. Thus the energy flux is found to be 20 times larger in Sim. B and the axial transport $\sim\langle V_z\rangle$ is 8 times larger.

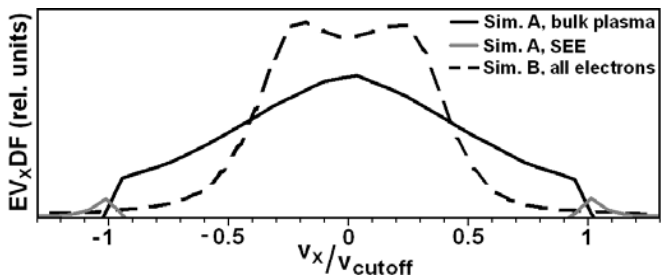

Fig. 3. $EV_xDF$ for electrons in the "middle" of the system (0.8cm < x < 1.7cm) at t = 10ms in both runs. $V_x$ is in units of $V_{cutoff} \approx 2.7 \times 10^8$ cm/s (the cutoff velocity for Sim. A with $\Phi \approx 21V$). For Sim. A, secondaries and bulk plasma electrons are plotted separately. In Sim. B, all electrons are secondaries. The "humps" in the $EV_xDF$ are responsible for the very strong two-stream fluctuations [20].

We have found with $B_x$ fixed at 100G, the inverse sheath tends to appear in simulations with electric field $E_z$ exceeding 200V/cm. The transition occurs because even the "coldest" electrons in the system have drift energies parallel to the wall oscillating from 0 to $2m_eV_D^2$. Therefore, when $V_D$ reaches a critical value, the average emission induced by secondary electron beams $\gamma_b$ will exceed unity. When this happens, a classical sheath, see Eq. (4), cannot maintain zero current. These results may have a connection to some important effects attributed to SEE in HT experiments. For wall materials with substantial SEE such as B.N.C, SEE effects become degrading at high voltages, leading to saturation of the temperature $T_e$ and maximum electric field [22]. In experiments, the discharge *voltage* is fixed. $E_z$ and $T_e$ are axially nonuniform, determined self-consistently with the axial transport and the balance between heating and losses. In EDIPIC, the *fields* are fixed and uniform, but the simulations suggest that as the voltage is increased in a HT, the E×B drift energy of electrons will reach a critical value in which the insulating sheaths begin to collapse. Further increases in $E_z$ and $T_e$ would be suppressed by the enhanced transport and energy loss.

The implications of these simulations are not limited to Hall thrusters. The E×B field just maintained the plasma temperature in the inverse sheath simulation; the plasma-wall interaction depended on the electron energies alone and was not device-dependent. Electrons can eject more than one secondary from many materials including insulators [7,19] and metals [7,9,23]. Conventional sheath theories may break down if the incident electrons eject more than one secondary on average. In this situation, ions do not "need" to be drawn to the wall. Instead, zero current can be maintained by an "inverse sheath," a positive surface charge shielded from the plasma by negative space charge at the interface. The wall potential is positive relative to the plasma and secondaries are pulled back to the wall to maintain zero current. The inverse sheath is fundamentally different from the "space-charge limited" sheath usually assumed to form at high temperatures in various systems [2,4,8,9,10]. The wall potential is still negative relative to the plasma in the SCL regime. Ions are drawn to the wall and most approaching electrons are reflected, as in a classical sheath. But in the inverse sheath regime, electrons travel unobstructed to the walls, causing extreme losses. This is important for plasma devices where the device performance is coupled to the plasma-wall interaction.